\documentclass{aa}

\usepackage{graphicx}
\usepackage{supertabular}
\usepackage{epsfig}
\usepackage{natbib}
\usepackage[varg]{txfonts}
\usepackage{hyperref}
\usepackage[normalem]{ulem}
\usepackage{multirow}
\usepackage{longtable}
\usepackage{pdflscape}
\usepackage{color,soul}
\usepackage{amssymb}

\newcommand{\logg}{\ensuremath{\log g}}

\newcommand{\feoh}{\ensuremath{[\mathrm{Fe/H}]}}

\newcommand{\Teff}{\ensuremath{T_{\mathrm{eff}}}}

\newcommand{\beq}{\begin{equation}}
\newcommand{\eeq}{\end{equation}}

\newcommand{\xtmean}[1]{\ensuremath{\left\langle #1\right\rangle}}

\newcommand{\MoH}{\ensuremath{\left[\mathrm{M}/\mathrm{H}\right]}}

\newcommand{\FeoH}{\ensuremath{\left[\mathrm{Fe}/\mathrm{H}\right]}}

\newcommand{\COBOLD}{{\tt CO$^5$BOLD}}
\newcommand{\LHD}{{\tt LHD}}

\newcommand{\ATLAS}{{\tt ATLAS9}}

\newcommand{\LINFOR}{{\tt Linfor3D}}
\newcommand{\MULTI}{{\tt MULTI}}

\definecolor{Blue}{rgb}{0,0.08,0.45}

\begin{document}

\title{Abundances of Mg and K in the atmospheres of turn-off stars in Galactic globular cluster 47~Tucanae\thanks{
Based on data obtained with the Very Large Telescope at the European Southern Observatory, programme ID: 081.D-0287(A).}}

\author{
   A.\,\v{C}erniauskas\inst{1}
   \and A.\,Ku\v{c}inskas\inst{1}
   \and J.\,Klevas\inst{1}
   \and V.\,Dobrovolskas\inst{1}
   \and S.\,Korotin\inst{2,3}
   \and P.\,Bonifacio\inst{4}
   \and H.-G.\,Ludwig\inst{5,4}
   \and E.\,Caffau\inst{4}
   \and M.\,Steffen\inst{6}
   }

\institute{
       Institute of Theoretical Physics and Astronomy, Vilnius University, Saul\.{e}tekio al. 3, Vilnius LT-10222, Lithuania \\
       \email{algimantas.cerniauskas@tfai.vu.lt}
           \and 
       Department of Astronomy and Astronomical Observatory, Odessa National University and Isaac Newton Institute of Chile Odessa
       branch, Shevchenko Park, 65014 Odessa, Ukraine
           \and
           Crimean Astrophysical Observatory, Nauchny 298409, Crimea
           \and
           GEPI, Observatoire de Paris, PSL Research University, CNRS, Place Jules Janssen, 92190 Meudon, France
       \and 
           Zentrum f\"ur Astronomie der Universit\"at Heidelberg, Landessternwarte, K\"onigstuhl 12, 69117 Heidelberg, Germany
           \and
           Leibniz-Institut f\"ur Astrophysik Potsdam, An der Sternwarte 16, D-14482 Potsdam, Germany
           }

\date{Received 27 July 2017 / Accepted 6 April 2018}

%============================================================================================================
\abstract
% context heading (optional) % {} leave it empty if necessary
{}
{We determined abundances of Mg and K in the atmospheres of 53 (Mg) and 75 (K) turn-off (TO) stars 
of the Galactic globular cluster 47~Tuc. The obtained abundances, together with those of Li, O, and 
Na that we had earlier determined for the same sample of stars, were used to search for possible relations 
between the abundances of K and other light elements, Li, O, Na, and Mg, as well as the connections between 
the chemical composition of TO stars and their kinematical properties.}
% methods  heading (mandatory)
{Abundances of Mg and K were determined using archival high resolution VLT~FLAMES/GIRAFFE spectra, in combination 
with the one-dimensional (1D) non-local thermodynamic equilibrium (NLTE) spectral synthesis methodology. Spectral line profiles were 
computed with the \MULTI\ code, using 1D hydrostatic \ATLAS\ stellar model atmospheres. We also utilized three-dimensional (3D) 
hydrodynamical \COBOLD\  and 1D hydrostatic \LHD\ model atmospheres for computing 3D--1D~LTE abundance corrections 
for the spectral lines of Mg and K, in order to assess the influence of convection on their formation in the 
atmospheres of TO stars.}
% results heading (mandatory)
{The determined average abundance-to-iron ratios and their root mean square (RMS) variations due to star-to-star abundance 
spreads were $\langle{\rm[Mg/Fe]}\rangle^{\rm 1D~NLTE}=0.47\pm0.12$, and $\langle{\rm [K/Fe]}\rangle^{\rm 1D~NLTE}=0.39\pm0.09$.
Although the data suggest the possible existence of a weak correlation in the [K/Fe]--[Na/Fe] plane, its statistical 
significance is low. No statistically significant relations between the abundance of K and other light elements 
were detected. Also, we did not find any significant correlations or anti-correlations between the [Mg/Fe] and [K/Fe] 
ratios and projected distance from the cluster center. Similarly, no relations between the absolute radial velocities 
of individual stars and abundances of Mg and K in their atmospheres were detected. The 3D--1D abundance corrections 
were found to be small ($\leq 0.1$\,dex) for the lines of Mg and K used in this study, thus indicating that the influence 
of convection on their formation is small.}
% conclusions heading (optional),  leave it empty if necessary
{}
%============================================================================================================
\authorrunning{\v Cerniauskas et al. }
\titlerunning{Abundances of Mg and K in TO of 47 Tuc}

\maketitle

%%%%%%%%%%%%%%%%%%%%%%%%%%%%%%%%%%%%%%%%
\section{Introduction}
%%%%%%%%%%%%%%%%%%%%%%%%%%%%%%%%%%%%%%%%

During the last decade studies of Galactic globular clusters (GGCs) opened a new chapter 
when it was discovered that GGCs may consist of multiple stellar generations. The first strong 
evidence in favor of this paradigm came from spectroscopic observations, which lead to the 
discovery of large star-to-star variation in the light element abundances within a given GGC 
\citep{K94,GSC04}, and, then, to the detection of various (anti-)correlations between the 
abundances of these elements, such as Na--O \citep{CBG09a} and Li--O \citep{PBM05,Shen10} 
correlations, and Na--Li \citep{BPM07} and Mg--Al \citep{CBG09a} anti-correlations. 
It is worth mentioning that these abundance trends are not seen in Galactic halo field stars. 
Further photometric observations have revealed the existence of multiple subsequences in the cluster 
color-magnitude diagrams (CMDs), all the way from the main sequence (MS) and up to the tip of 
the red giant branch (RGB) \citep{PBA07,MPB12}. All these findings suggest that 
stars in the GGCs may have formed during two or more star formation episodes \citep[see, e.g.,][]{GVL12}, 
thus contradicting the earlier notion that GGCs are perfect examples of simple stellar populations.

The most popular theories are that either massive asymptotic giant branch (AGB) stars \citep[e.g.,][]{VD01} or fast-rotating massive stars \citep[e.g.,][]{DMC07} 
could have enriched the second-generation stars in Na and Al, and depleted 
them in O and Mg. Other scenarios, such as enrichment by binary stars \citep{dPLI09} and 
early disk accretion \citep{BLd13} have been discussed, too. However, none of them can explain, for example, all observed abundance (anti-)correlations simultaneously 
\citep[see discussion in, e.g.,][]{BCS15}. From the theoretical point of view, new ideas 
regarding the possible polluters are needed. From the observational side, it would be 
desirable to identify new chemical elements that would allow us to discern between the different already 
proposed self-enrichment scenarios of the GGCs or, possibly, help to suggest new ones.

Potentially, new clues in this context may come from the investigations of potassium abundance. 
Since K is synthesized mainly via oxygen burning in high-mass stars, it is unlikely 
that the atmospheric K abundance would undergo any appreciable changes during the course of the stellar 
evolution of the low-mass stars in the GGCs. \citet{MBI12} presented an analysis of RGB stars in NGC 2419 
($\FeoH=-2.09$) in which the first hints of K--Mg anti-correlation and bimodal distribution of [Mg/Fe] 
ratio were detected. The authors also confirmed an unusually large ($\approx2$\,dex) spread and depletion 
(to $\approx \rm-1$ dex) in the magnesium abundance. A more recent study of NGC~2808 ($\FeoH=-1.1$) 
by \citet[][]{MBM15} has revealed the existence of statistically significant correlations of the [K/Fe] 
abundance ratio with [Na/Fe] and [Al/Fe], and anti--correlations with [O/Fe] and [Mg/Fe]. A fraction of the stars in both clusters is strongly enhanced in helium, with $Y=0.34$ in NGC~2808 
\citep{MMP14} and $Y=0.42$ in NGC~2419 \citep{dDM11}. These values are much higher than those 
observed in other GGCs \citep{MMD14}. \citet{MBM15} suggested a self-enrichment model proposed by 
\citet{DDC12} as one of the possible explanations of $\rm [K/Fe] - [Mg/Fe]$ anti-correlation. In this 
scenario, Mg-poor/K-rich (extreme population) stars formed from the ejecta of AGB and super-AGB stars. 
However, this scenario still requires some fine tuning in the nuclear reaction cross-sections and burning 
temperatures to explain the observed trends satisfactorily. 

In their recent study of 144 RGB stars in 47~Tuc, \citet{MMB17} claimed a detection of a mild 
$\rm [K/Fe] - [Na/Fe]$ correlation and $\rm [K/Fe] - [O/Fe]$ anti-correlation. This is in contrast 
with the results of our recent study of Na, Mg, and K abundances in the atmospheres of RGB stars in 
47~Tuc, where we found no statistically significant relations between either the abundances of 
different elements, or the abundances and kinematical properties of the cluster stars \citep[hereafter 
Paper I]{CKK17}. It is conceivable, however, that non-detection in our case could be due to a smaller 
sample of RGB stars studied: 32 in our work versus 144 in the study of \citet{MMB17}. Apart from 
these two studies, the only other investigation of K abundances in 47~Tuc was done by \citet{CGB13} 
who determined K abundances in three turn-off (TO) and nine subgiant branch (SGB) stars. Obviously, the sample size 
used in the latter study was too small to search for possible relations between the elemental 
abundances among little-evolved stars.

Given this somewhat ambiguous situation, investigation of K abundances in the atmospheres of TO stars 
in 47~Tuc could be very interesting, especially if based on a larger sample of stars 
than that used by \citet{CGB13}. It is well know that the cores of unevolved low-mass stars do not reach 
temperatures high enough for $\rm Ne-Na$ and/or $\rm Mg-Al$ cycles to operate. Moreover, their convective 
envelopes are not deep enough to bring up to the surface the products of proton-capture reactions. 
Therefore, the atmospheres of these stars should have retained their primordial chemical composition,
 unless their atmospheres have been contaminated by accreted chemical elements synthesized 
in other stars. In the present study we therefore determine abundances of Mg and K in the atmospheres 
of 53 and 75 (TO) stars in 47~Tuc, respectively (abundances of both elements were 
obtained in 44 stars).
We then use this data to search for possible relations between the abundances of K 
and Mg, and those of other light elements, Li, O, and Na (with their abundances taken from \citealt{DKB14}), 
as well as relations between the elemental abundances and kinematical properties of TO stars.
 
The paper is structured as follows. In Sect.~\ref{sect:method} we present spectroscopic data used in 
our study and outline the procedures used to determine abundances of Mg and K using one-dimensional (1D) non-local 
thermodynamic equilibrium (NLTE) methodology. Analysis and discussion of the obtained results is 
presented in Sect.~\ref{sect:discuss}, while the main findings of this paper and final conclusions 
are outlined in Sect.~\ref{sect:conclus}.

%%%%%%%%%%%%%%%%%%%%%%%%%%%%%%%%%%%%%%%%
\section{Methodology}\label{sect:method}
%%%%%%%%%%%%%%%%%%%%%%%%%%%%%%%%%%%%%%%%

Abundances of Mg and K were determined using 1D hydrostatic \ATLAS\ model atmospheres and 
1D~NLTE abundance analysis methodology. Additionally, we also utilized three-dimensional (3D) hydrodynamical \COBOLD\ 
and 1D hydrostatic \LHD\ model atmospheres to compute the 3D--1D abundance corrections using the assumption of local thermodynamic equilibrium 
(LTE; see Sect.~\ref{sect:abn_corr} for details). This was done in order to study the importance of convection 
in the formation of the spectral lines of \ion{Mg}{i} and \ion{K}{i} used in this study, though the obtained 
corrections were not applied to determine 3D-corrected elemental abundances (see Sect.~\ref{sect:abn_corr} 
for details). A brief description of all steps involved in the abundance analysis is provided below.

%===================================================
\subsection{Spectroscopic data\label{sect:obs_data}}

In this work we used the same sample of TO stars of 47~Tuc as in \citet{DLG10} and \citet{DKB14}. 
We note that in the two previous studies abundances of Mg and K in the atmospheres 
of these stars were not determined.

In the abundance analysis we utilized high-resolution archival VLT~FLAMES/GIRAFFE spectra of TO 
stars that were reduced by and used in \citet[][programme ID: 081.D-0287(A), PI: Shen]{DKB14}. The spectra 
were obtained in Medusa mode using HR~18 setup (746.8 -- 788.9\,nm, \textit{R}=18400). In total, 
116 fibers were dedicated to target stars and 16 were used for sky spectra. The continuum 
normalization procedure was completed using IRAF \citep{T86} \textit{continuum} task 
\citep[see][for details]{DKB14}.

Effective temperatures and surface gravities of the sample TO stars were taken from 
\citet{DKB14}. The former were determined by fitting the wings of H$\alpha$ line profiles, 
while the latter were obtained using the classical relation between the surface gravity, 
mass, effective temperature, and luminosity.

%================================================
\subsection{Model atmospheres\label{sect:mod_atm}}

We used two types of 1D hydrostatic model atmospheres, \ATLAS\ and \LHD, and 3D hydrodynamical \COBOLD\ model atmospheres:

\begin{list}{--}{}

\item \ATLAS: for the atmospheric parameters of each individual sample star we computed a 1D 
hydrostatic model atmosphere using the \ATLAS\ code \citep{K93,S05}. The model atmospheres 
were calculated using ODFNEW $\MoH=-1.0$ opacity tables \citep[][]{CK03}, with the $\alpha-$element 
enhancement of $[\alpha/{\rm Fe}]=+0.4$. The mixing length parameter was set to $\alpha_{\rm MLT}=1.25$ 
and the overshooting was switched off. These model atmospheres 
were used in the 1D NLTE abundance analysis of Mg and K for synthesizing spectral line profiles 
with the \MULTI\ code (see Sect. \ref{sect:1Dabund});

\item  \LHD: the 1D hydrostatic \LHD\  models were computed using the \LHD\ 
model atmospheres code \citep{CLS08}, which utilises chemical composition, equation of state, 
and opacities identical to those used in the 3D hydrodynamical \COBOLD\ model atmospheres 
(see below). In order to compute the 3D--1D abundance corrections for magnesium and potassium 
(see Sect. \ref{sect:abn_corr}), atmospheric parameters of the \LHD\ models were matched to 
those of the \COBOLD\ model atmospheres;

\item \COBOLD: the 3D hydrodynamical \COBOLD\ model atmosphere code solves time-dependent 
equations of hydrodynamics and radiation transfer on a Cartesian grid  \citep[][]{FSL12}. 
We used four 3D hydrodynamical model atmospheres from the CIFIST grid 
\citep{LCS09}, which we utilised for computing the 3D--1D abundance corrections. Atmospheric 
parameters of the \COBOLD\ models are provided in Table~\ref{tab:3Dmodels2}. The \COBOLD\ 
and \LHD\ models were computed using an identical chemical composition, equation of state, 
opacities, and radiative transfer scheme.

\end{list}

\begin{table}[tb]
\caption{Parameters of the 3D hydrodynamical \COBOLD\ model atmospheres used in this work. \label{tab:3Dmodels2}}  
\centering                 
\setlength{\tabcolsep}{3pt}
\begin{tabular}{c c r@{}l c c}     
\hline\hline                   
  \Teff, K & \logg\ & \multicolumn{2}{c}{$\MoH$} & Grid dimension, Mm & Grid resolution\\    
           &        &       && x $\times$ y $\times$ z & x $\times$ y $\times$ z\\
\hline                      
 5470 & 4.0 &   &$0.0$    & $20.3\times20.3\times10.6$ & $140\times140\times150$ \\    
 5530 & 4.0 &  -&$1.0$    & $19.9\times19.9\times10.6$ & $140\times140\times150$ \\
 5930 & 4.0 &   &$0.0$    & $25.8\times25.8\times12.5$ & $140\times140\times150$ \\
 5850 & 4.0 &  -&$1.0$    & $25.8\times25.8\times12.5$ & $140\times140\times150$ \\
\hline                     
\end{tabular}
\end{table}

%================================================================================
\subsection{Determination of 1D~NLTE abundances of Mg and K \label{sect:1Dabund}}

The 1D~NLTE abundances of Mg and K were determined using 
the \MULTI\ code \citep[][]{C86} modified by \citet[][]{KAL99}. The code computes theoretical 
line profiles using 1D \ATLAS\ model atmospheres and a model atom of a given chemical element 
(Sect.~\ref{sect:atom}). Atomic parameters of the spectral lines that were used in our study are 
provided in Table~\ref{elem}. In the case of \ion{Mg}{}, line parameters were taken from the Vienna atomic line database (VALD-3) 
\st{atomic} database \citep[][]{PKRWJ95,KD11}. For K, $\log gf$ value was taken from \citet{M91} while 
the line broadening constants are from VALD-3. We stress that in the case of K, the NLTE approach in 
the abundance analysis is critical since for this element 1D~NLTE--LTE abundance corrections 
are typically very large, reaching from $-0.5$ to $-0.7$\,dex \citep{TZC02}. Besides, they 
tend to be larger for the late-type stars and increase with decreasing metallicity \citep{ASK10}.
   
\begin{figure}[t!]
\begin{center}
\includegraphics[scale=1.40]{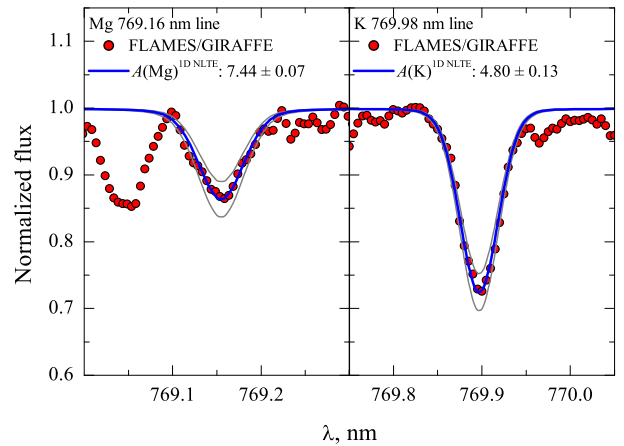}
\caption{Typical fits of synthetic \ion{Mg}{} and \ion{K}{} line profiles (solid blue lines) to those in the 
observed GIRAFFE spectrum (filled red circles) of the target TO star 47Tuc45982 ($\Teff=5707$ K, $\log g=4.00$). 
We also provide the abundances determined from each observed line, $A{\rm (X)}$, together with their errors 
(see Sec.~\ref{sect:abund_err}). Thin gray lines show synthetic line profiles computed with the abundances 
altered by $\pm 0.2$\,dex.}
\label{fig:Kline}
\end{center}
\end{figure}

Abundances of each element were determined by fitting theoretical line 
profiles to those observed in a given TO star. A typical example of the obtained 
best fit is shown in Fig. \ref{fig:Kline}. During the fitting procedure, we used a fixed microturbulence velocity of 1.0\,km~s$^{-1}$ for all sample stars, 
while the macroturbulence velocity was varied during each iteration as a free parameter to obtain the 
best match to the observed line profile. The macroturbulence velocities of stars in our 
sample were in the range of 1 to 5 km~s$^{-1}$. A fixed value of $\FeoH^{\rm 1D~LTE}=-0.76$ 
from \citet[][]{CBG09a} was used throughout this study.

\begin{table}
\caption{Atomic parameters of the spectral lines used in the abundance determinations 
of Mg and K. Natural ($\gamma_{\rm{rad}}$), Stark ($\frac{\gamma_4}{N_{\rm{e}}}$), and van der 
Waals ($\frac{\gamma_6}{N_{\rm{H}}}$) broadening constants computed using classical 
prescription are provided in the last three columns.}
\label{param}
\vspace{-5mm}
\begin{center}
\scalebox{0.95}{
\setlength{\tabcolsep}{2pt}
\begin{tabular}{lcccccc}
\hline
\hline
Element & $\lambda$, nm & $\chi$, eV & log$\textit{gf}$ & log$\gamma_{\rm{rad}}$ & log$\frac{\gamma_4}{N_{\rm{e}}}$ & log$\frac{\gamma_6}{N_{\rm{H}}}$ \\
\hline
Mg I & 769.16 & 5.753 & $-0.78$ & 7.57 & $-3.25$ & $-6.83$ \\
K  I & 769.89 & 0.000 & $-0.17$ & 7.56 & $-5.44$ & $-7.45$ \\
\hline
\end{tabular}
\label{elem}
}
\end{center}
\end{table}

%=======================================================
\subsubsection{Model atoms of Mg and K \label{sect:atom}}

The model atoms of \ion{Mg}{} and \ion{K}{} that were used in our study are briefly described below;  
for more details see Paper~I and references therein.
In the case of Mg, we used the model atom from \citet[][]{MSK04}. It consisted of 84 levels of \ion{Mg}{i}, 
12 levels of \ion{Mg}{ii}, and the ground state of \ion{Mg}{iii}. In the computation of departure 
coefficients, radiative transitions between the first 59 levels of \ion{Mg}{i} and ground level of 
\ion{Mg}{ii} were taken into account. 

The model atom of \ion{K}{} was taken from \citet[][]{ASK10} and consisted of 20 levels of \ion{K}{i} 
and the ground level of \ion{K}{ii}. In addition, another 15 levels of \ion{K}{i} and seven levels of 
\ion{K}{ii} were used to ensure particle number conservation. The total number of bound-bound 
radiative transitions taken into account was 62 \citep[see] [for further details]{ASK10}.

%==================================================================
\subsubsection{ 1D~NLTE abundances of Mg and K in the atmospheres of TO stars in 47 Tuc \label{sect:MgK47Tuc}}

Before the determination of Mg and K abundances, \ion{Mg}{i} and \ion{K}{i} lines in the spectra of all stars studied were carefully 
inspected for blends and/or possible contamination by telluric lines (to remind, we had only one spectral line per element available for the abundance determination in the spectrum of each TO star). This 
inspection revealed significant star-to-star variation in terms of the line quality.
In order to take this into account, 
we grouped \ion{Mg}{i} and \ion{K}{i} lines into three 
classes according to their quality, the latter determined by visual inspection using the following criteria:

 \begin{list}{--}{}
        \item A-class: strong or moderately strong lines with well-resolved line profiles;
        \item B-class: lines that are moderately blended with telluric lines, or lines that were insufficiently 
                                   resolved in the line wings;
        \item C-class: lines with weak and/or poorly defined profiles, or significantly blended lines.
 \end{list}

\begin{figure*}[!t]
        \centering
        \mbox{\includegraphics[width=17cm]{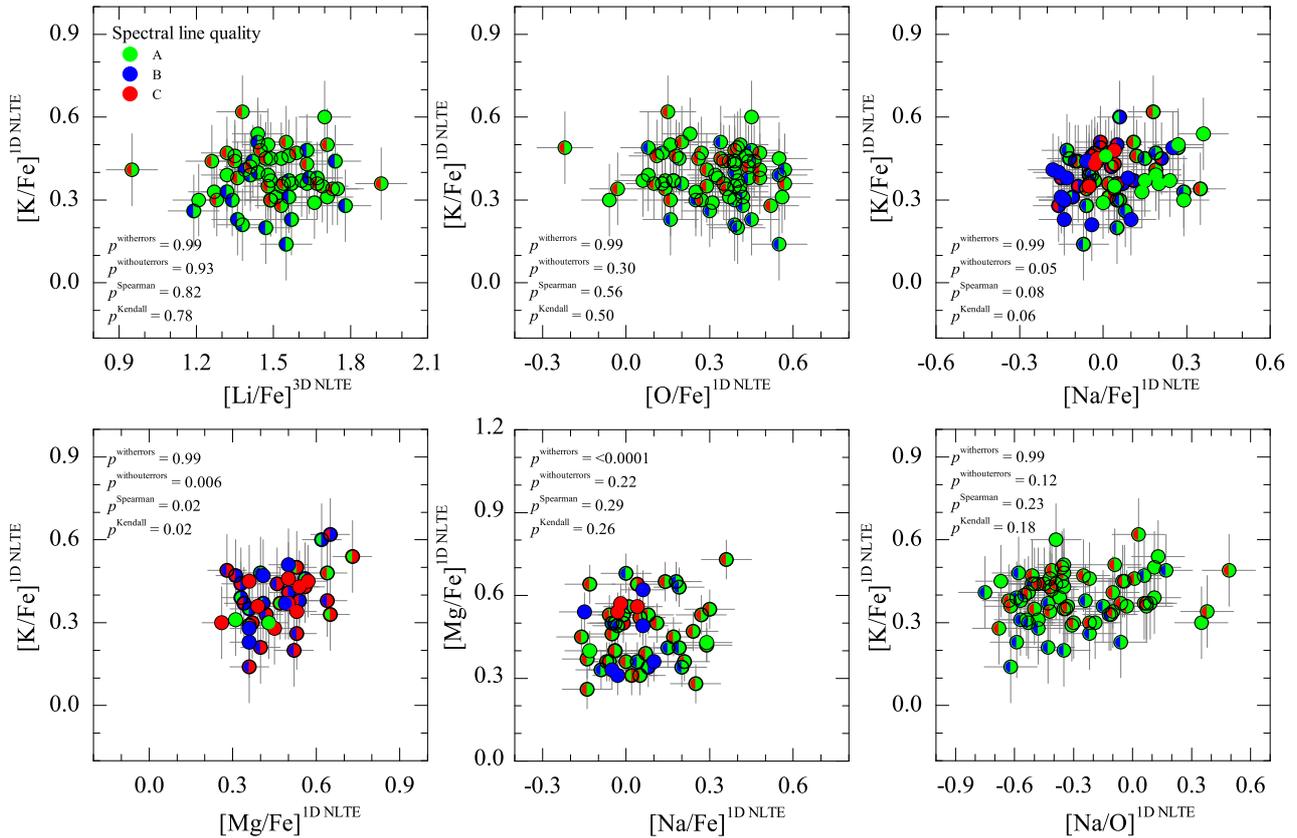}}
        \caption{Abundance-to-iron ratios of Li, O, Na, Mg, and K in the sample of TO stars,  
        shown in various abundance-abundance planes. Colors on the left and right sides 
        of the symbols correspond to the quality (symbols denoting different quality classes are shown in the top-left panel) of spectral lines used to determine 
        abundances of the light elements in a given star plotted on the $y$ and $x$ axes, respectively (see Sect. \ref{sect:MgK47Tuc} 
        for details). The values of two-tailed probabilities, $p$, computed using Pearson's 
        parametric correlation coefficients (with and without abundance errors, $p^{\rm witherrors}$ 
        and $p^{\rm withouterrors}$, respectively), Spearman's, and Kendall's non-parametric rank-order 
        correlation coefficients are given in the corresponding panels (see Sect.~\ref{sect:discuss_relat}).}
        \label{fig:abund-ratios}
\end{figure*}

\noindent These quality flags are marked by different colors in Fig.~\ref{fig:abund-ratios}. 
The line of \ion{Mg}{i} and the line of \ion{K}{i} in the spectrum of each star were always assigned to their individual quality classes. As a consequence, 
even in the spectrum of the same star, \ion{Mg}{i} and \ion{K}{i} lines could belong to different quality classes, for example, Mg line to A-class, K line to C-class. 
This explains why both Mg and K could not be determined in all the stars, because for some stars only one of the two lines was suitable for abundance determination.

To verify that the spectral lines used in our study are not seriously 
affected by telluric lines, such as telluric A band in the vicinity of the \ion{K}{i} 
769.89\,nm line, we used (a) telluric lines identified in the spectrum 
of fast-rotating O6.5~III spectral type star HD94963, which was taken from the UVES POP spectral library \citep{BJL03}; and (b) a synthetic spectrum of the 
atmospheric transmission computed with the {\tt TAPAS} tool \citep{BLF14} 
for the dates when the observations were done.
 
\begin{figure}[!t]
        \centering
        \includegraphics[width=9cm]{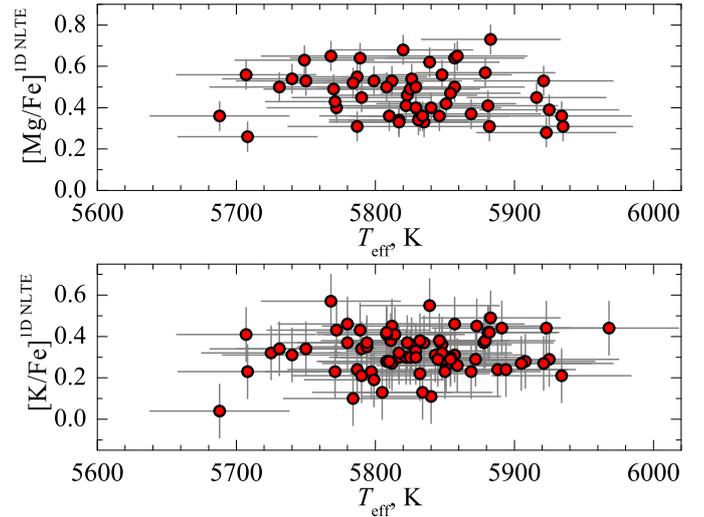}
        \caption{[Mg/Fe] (top) and [K/Fe] (bottom) abundance ratios 
        determined in the sample of TO stars in 47~Tuc and plotted versus the effective 
        temperature of individual stars. Error bars show uncertainties that were 
        computed as described in Sect.\ref{sect:abund_err}.}
        \label{fig:abund-teff}
\end{figure}

The 1D~NLTE abundances of Mg and K were then determined by fitting synthetic spectral line profiles 
to those observed in the spectra of TO stars. We verified that the determined abundances show no 
dependence on the effective temperature (Fig.~\ref{fig:abund-teff}).

The iron abundance among the stars in 47 Tuc is very homogeneous: from 147 stars observed with 
GIRAFFE \citet{CBG09a} found $\FeoH=-0.743 \pm\,0.003$\,(stat) $\pm\,0.026$\,(syst) and from 11 stars observed with 
UVES, \citet{CBG09b} determined $\FeoH=-0.768 \pm\,0.016$\,(stat) $\pm\,0.031$\,(syst), in both cases with the 
assumed iron abundance $\textit{A}(\rm Fe)=7.54$ from \citet{GCC03}. In our GIRAFFE spectra we can measure 
about 20 Fe I lines, which are of poor quality, due to the low signal-to-noise ratios, $S/N$, of the spectra and contamination 
from telluric lines. As a consequence, the line to line scatter is 0.2 dex or larger for our 
stars. We therefore consider it more robust to assume for each star the mean Fe abundance of the cluster 
that we take as the average of the measurements from the UVES spectra by \citet{CBG09b} and the GIRAFFE 
spectra by \citet{CBG09a}: $\FeoH=-0.76$. Solar Mg and K abundances, $\textit{A}(\rm Mg)^{1D~NLTE}_{\odot}=7.64 \pm 0.05$ and $\textit{A}(\rm K)^{1D~NLTE}_{\odot}=5.10 \pm 0.07$, were
determined in this work (see Appendix~\ref{sect_app:ArcSun}). The 1D~NLTE abundance ratios of [Li/Fe], [O/Fe], and 
[Na/Fe] were taken from \citet{DKB14} where authors determined them using the same [Fe/H] value as utilized in the present study. More information about the procedure of Li, O, and Na abundance determination 
see Sect.~5.2.2, 5.2.3, and 5.3 in \citet{DKB14}. 

The determined average element-to-iron abundance ratios in the sample of TO stars are
$\langle{\rm[Mg/Fe]}\rangle^{\rm 1D~NLTE} = 0.47\pm0.12$ (53 objects) and $\langle{\rm[K/Fe]}\rangle^{\rm 1D~NLTE} = 0.39\pm0.09$ 
(75 objects; numbers after the $\pm$ sign are RMS abundance variations due to star-to-star scatter). 
In Fig. \ref{fig:abund-ratios} we show [K/Fe], [Mg/Fe], [Na/Fe], [O/Fe], and [Li/Fe] abundance ratios 
plotted in various abundance-abundance planes. 

%=======================================================================
\subsubsection{3D--1D abundance corrections for Mg and K \label{sect:abn_corr}}

To assess the role of convection in the formation of Mg and K lines in the atmospheres of TO 
stars, we used 3D hydrodynamical \COBOLD\ and 1D hydrostatic \LHD\ model atmospheres (Sect. 
\ref{sect:mod_atm}). The two types of model atmospheres were used to compute 3D--1D LTE abundance 
corrections for the spectral lines of \ion{Mg}{i} and \ion{K}{i} utilized in our study (corrections 
for the lines of \ion{O}{i} and \ion{Na}{i} were computed earlier in \citealt{DKB14}). Spectral 
line synthesis computations were carried out with the \LINFOR\ spectral synthesis package.\footnote{\url{http://www.aip.de/Members/msteffen/linfor3d}.} 

The procedure used to compute the 3D--1D~LTE abundance corrections, $\Delta_{\rm 3D-1D~LTE}$, 
was identical to that utilized in Paper~I. Since the abundance corrections of Mg and K showed 
little variation with the spectral line strength, in case of each element they were computed 
for two values of line equivalent width, $W$ (corresponding to ``weak'' and ``strong'' spectral lines) 
that bracketed the range measured in the observed spectra of the sample TO stars. Equivalent widths 
used for the weakest lines were 4\,pm and 15\,pm, while for the strongest lines we used 7\,pm 
and 18\,pm, in the case of Mg, and K, respectively. Microturbulence velocity in the 3D model 
atmosphere was determined by applying Method~1 described in \citet[][]{SCL13} and was subsequently 
used in the spectral line synthesis with the \LHD\ model atmospheres (see Paper~I for details). 

The obtained 3D--1D~LTE abundance corrections, $\Delta_{\rm 3D-1D~LTE}$, are provided in Table~\ref{tab:abund_corr}. 
For both lines, they do not exceed $0.09$\,dex, which allows us to conclude that the influence 
of convection on the formation of \ion{Mg}{i} and \ion{K}{i} lines in the atmospheres of TO stars is minor. 

\begin{table}
        \centering
        \caption{\st{Obtained} 3D--1D abundance corrections, $\Delta_\mathrm{3D-1D~LTE}$, computed for 
        different strengths of \ion{Mg}{i} and \ion{K}{i} lines used in 
        this work (see text for details).}
        \label{tab:abund_corr}
        \centering
        \begin{tabular}{lccccc}
                \hline\hline 
                Element      & $\lambda_{\rm central}$ & \multicolumn{3}{c}{$\Delta_\mathrm{3D-1D~LTE}$, dex} \\
                             &    nm                   & weak    & strong                         \\
                \hline
                \ion{Mg}{i}  & 769.16 nm               & $+0.04 $  & $+0.05 $ \\
                \ion{K}{i}   & 769.89 nm                   & $-0.09$  & $-0.03$  \\
                \hline 
        \end{tabular}
\end{table}

We note that the obtained abundance corrections were not used for obtaining 3D-corrected abundances which, in principle, could be done by adding the 3D--1D~LTE corrections to the determined 1D~NLTE abundances of Mg and K. Such 
a procedure was avoided for two reasons. First, abundances obtained in this way would be generally 
different from those that would be obtained using the full 3D~NLTE approach  \citep[see, e.g.,][]{KKS16}. 
Second, the determined 3D--1D abundance corrections are small, therefore applying them would result in 
a small and nearly uniform shift of the abundances determined in all our sample stars. In fact, only 
in the case of K is the difference in the 3D--1D abundance corrections obtained for weak and strong lines somewhat 
larger, $0.06$\,dex, while for Mg this difference is only $0.01$\,dex. 
Our tests have shown that if these (small) 3D--1D abundance corrections are taken into account, 
our conclusions regarding the intrinsic abundance spreads and possible existence of different relations 
between, for example, the abundance of K and those of other light elements, remain unaltered (see Sect.~\ref{sect:discuss}).

%=================================================================================
\subsubsection{Uncertainties in the determined abundances of Li, O, Na, Mg, and K} \label{sect:abund_err}

The uncertainties in the determined abundances occur for two main reasons: (i) inaccurate 
determination of atmospheric parameters (effective temperature, surface gravity, and microturbulence 
velocity); and (ii) uncertainties in the spectral line profile fitting (due to the choice of continuum 
level and inaccurate spectral line profile fit). Individual contributions to the total uncertainty in the determined abundances of O, Na, Mg, and K arising from the different error sources
were estimated in the following way:\footnote{An identical error determination procedure was also used in Paper~I. In this procedure errors due to uncertainties in the atomic line parameters, as well as various systematic errors, were ignored. Apart from the uncertainties for Mg and K, for consistency we also re-derived the errors for Li, O, and Na, which had their abundances determined in \citet{DKB14}.}

\begin{table}[t!]
        \begin{center}
        \caption{Errors in the abundances of Li, O, Na, Mg, and K determined in TO stars of 47~Tuc. 
        The sign $\pm$ or $\mp$ reflects the change in the elemental abundance, which occurs due to 
        the increase (top sign) or decrease (bottom sign) by the value of typical uncertainty (Sect.~\ref{sect:abund_err}) in \Teff, $\log g$, $\xi_{\rm t}$, continuum placement, and the line profile fit (cols.~4--8). For 
        example, an increase in the effective temperature leads to increasing abundance ($\pm$), 
        while increasing microturbulence velocity results in decreasing abundance ($\mp$). The total estimated uncertainty is provided in col.~9.}
        \vspace{-5mm}
                \resizebox{\columnwidth}{!}{            
                \begin{tabular}{lccccccccc}
                        \hline\hline
                        Element & Line & Line     &  $\sigma(\Teff)$ & $\sigma(\log g)$  &      $\sigma(\xi_{\rm t})$    &  $\sigma(\rm cont)$ &  $\sigma(\rm fit)$  & $\sigma(A)_{\rm tot}$  \\ [0.5ex]
                        & $\lambda$, nm          & quality  &             dex              &             dex           &        dex         &   dex    &    dex    &   dex             \\
                        \hline
                        \ion{Li}{i}   & 670.80 & A & $\pm0.09$ & $\mp0.01$  & $\mp0.01$ & $0.03$  & $0.01$ &  0.10 \\
                        &                      & B & $\pm0.09$ & $\mp0.01$  & $\mp0.01$ & $0.03$  & $0.02$ &  0.10 \\
                        &                      & C & $\pm0.09$ & $\mp0.01$  & $\mp0.01$ & $0.03$  & $0.03$ &  0.10 \\
                        \ion{O}{i}    & 777.19 & A & $\pm0.09$ & $\mp0.03$  & $\mp0.02$ & $0.02$  & $0.01$ &  0.10 \\
                        &                      & B & $\pm0.09$ & $\mp0.03$  & $\mp0.02$ & $0.02$  & $0.02$ &  0.10 \\
                        &                      & C & $\pm0.09$ & $\mp0.03$  & $\mp0.02$ & $0.02$  & $0.03$ &  0.10 \\
                        \ion{O}{i}    & 777.53 & A & $\pm0.09$ & $\mp0.03$  & $\mp0.02$ & $0.02$  & $0.01$ &  0.10 \\
                        &                      & B & $\pm0.09$ & $\mp0.03$  & $\mp0.02$ & $0.02$  & $0.03$ &  0.10 \\
                        &                      & C & $\pm0.09$ & $\mp0.03$  & $\mp0.02$ & $0.02$  & $0.04$ &  0.11 \\
                        \ion{Na}{i}   & 818.32 & A & $\pm0.06$ & $\mp0.02$  & $\mp0.06$ & $0.02$  & $0.01$ &  0.09 \\
                        &                      & B & $\pm0.06$ & $\mp0.02$  & $\mp0.06$ & $0.02$  & $0.02$ &  0.09 \\
                        &                      & C & $\pm0.06$ & $\mp0.02$  & $\mp0.06$ & $0.02$  & $0.03$ &  0.09 \\
                        \ion{Na}{i}   & 819.48 & A & $\pm0.06$ & $\mp0.02$  & $\mp0.06$ & $0.03$  & $0.01$ &  0.09 \\
                        &                      & B & $\pm0.06$ & $\mp0.02$  & $\mp0.06$ & $0.03$  & $0.02$ &  0.09 \\
                        &                      & C & $\pm0.06$ & $\mp0.02$  & $\mp0.06$ & $0.03$  & $0.03$ &  0.10 \\
                        \ion{Mg}{i}   & 769.16 & A & $\pm0.04$ & $\mp0.03$  & $\mp0.04$ & $0.03$  & $0.01$ &  0.07 \\
                        &                      & B & $\pm0.04$ & $\mp0.03$  & $\mp0.04$ & $0.03$  & $0.01$ &  0.07 \\
                        &                      & C & $\pm0.04$ & $\mp0.03$  & $\mp0.04$ & $0.03$  & $0.02$ &  0.07 \\
                        \ion{K}{i}    & 769.89 & A & $\pm0.08$ & $\mp0.02$  & $\mp0.10$ & $0.02$  & $0.01$ &  0.13 \\
                        &                      & B & $\pm0.08$ & $\mp0.02$  & $\mp0.10$ & $0.02$  & $0.02$ &  0.13 \\
                        &                      & C & $\pm0.08$ & $\mp0.02$  & $\mp0.10$ & $0.02$  & $0.03$ &  0.13 \\ 
                        \hline
                \end{tabular}}
            \label{tab_app:abund_err}
        \end{center}
        
        \vspace{-5mm}
\end{table}

\begin{figure}[tb]
        \centering
        \mbox{\includegraphics[width=8cm]{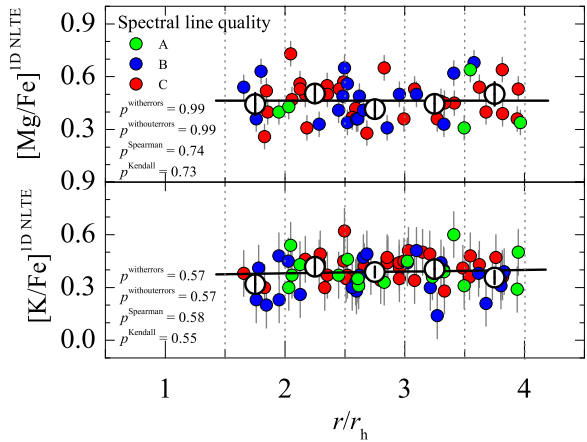}}
        \caption{Abundance-to-iron ratios of Mg and K 
        plotted versus the projected distance from the cluster center, $r/r_{\rm h}$ (small filled 
        circles; $r_{\rm h}$ is a half-mass radius of 47~Tuc, $r_{\rm h}=174^{\prime\prime}$, 
        taken from \citealt{TDK93}). Symbol color denotes quality (class) of the spectral 
        lines from which the abundance was determined. 
        Large open circles are average abundance ratios
        computed in non-overlapping $\Delta r/r_{\rm h}=1$ wide distance bins (marked by 
        the vertical dashed lines; RMS scatter of the abundance 
        ratios in a given bin is shown by the black vertical error bars). Black solid lines are 
        linear fits to the data of individual stars, with the $p$-values obtained using different tests (see text) marked 
        in the corresponding panels.}
        \label{fig:abund-radial}
\end{figure}

\begin{figure}[tb]
        \centering
        \mbox{\includegraphics[width=8cm]{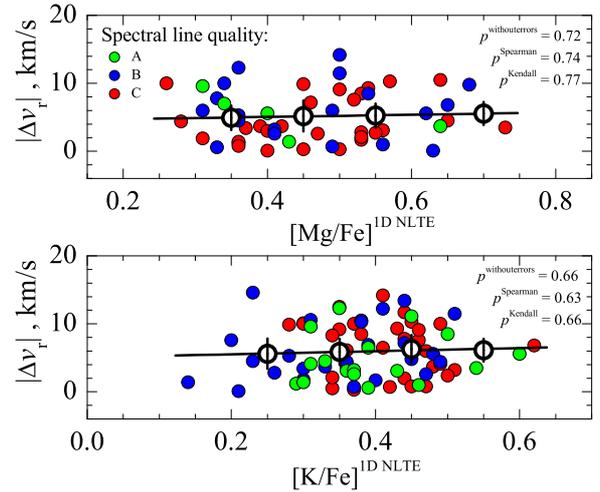}}
        \caption{Absolute radial velocities of TO stars in 47~Tuc, $\left| \Delta v_{\rm r}\right|$, 
        plotted versus [Mg/Fe] and [K/Fe] abundance ratios measured in their atmospheres. 
        Other symbols and notations are as in Fig.~\ref{fig:abund-radial}.}
        \label{fig:disper-radial}
\end{figure}

\begin{list}{--}{}

\item $\sigma(\Teff)$ (Table~\ref{tab_app:abund_err}, col.~4): effective temperatures that were used in our study were obtained in \citet{DKB14} using the H$\alpha$ line profile fitting. The authors estimated that the uncertainty in their determined effective temperatures was $\pm100$\,K . We used this value to evaluate the influence of the uncertainty in \Teff\ on the determined elemental abundances, $\sigma(\Teff)$;

\item $\sigma(\log g)$ (Table~\ref{tab_app:abund_err}, col.~5): the error in the surface gravity, 
$\log g$, $\pm0.04$\,dex, was obtained from the individual components: error in the effective temperature 
($\pm100$\,K), error in luminosity ($\pm 0.03~{\rm L}_{\odot}$), estimated from the photometric error in 
$M_{\rm V}$, and stellar mass ($\pm0.01~{\rm M}_{\odot}$, obtained from the isochrones). However, the 
error in surface gravity obtained in this way was, in our view, unrealistically low. Therefore, to estimate 
the resulting errors in the determined abundances, $\sigma(\log g)$, a more conservative error of $\pm 0.1$\,dex was used;

\item $\sigma(\xi_{\rm t})$ (Table~\ref{tab_app:abund_err}, col.~6): the error in the microturbulence 
velocity was estimated by evaluating the slope uncertainty in the ${\rm [Fe/H]}-W$ plane ($W$ is line equivalent width). In this procedure we used six TO stars for which we were able to determine Fe abundance using individual \ion{Fe}{i} lines. 
The average slope error determined in this way was $\pm0.001$ dex/pm, which corresponds to the error in $\xi_{\rm t}$ of $\pm0.29$\,km/s. 
This value was used as a representative uncertainty in $\xi_{\rm t}$ to estimate the resulting abundance errors, $\sigma(\xi_{\rm t})$;

\item $\sigma(\rm cont)$ (Table~\ref{tab_app:abund_err}, col.~7): the error in continuum determination was 
estimated in the same way as in Paper~I, by measuring the dispersion at the continuum level in the 
spectral windows expected to be free of spectral lines, both of stellar and telluric origin (see Paper~I for details). The continuum was then shifted by the amount of this uncertainty to obtain the resulting error in the determined abundance, $\sigma(\rm cont)$;

\item $\sigma(\rm fit)$ (Table~\ref{tab_app:abund_err}, col.~8): to estimate the errors in the line profile 
fitting, we computed RMS deviation between the observed and synthetic line profiles, which were converted into 
the uncertainties in the line equivalent width, $W$, and, finally, into the errors in the 
determined abundances, $\sigma(\rm fit)$ (see Paper~I for details).  
\end{list}

Individual errors were added in quadratures to obtain the total error in the determined abundances 
of O, Na, Mg, and K (col. 9 in Table~\ref{tab_app:abund_err}). The obtained total errors were 
further used in Sect~\ref{sect:ML_test} to estimate the possible intrinsic spreads in the obtained elemental 
abundances (see Sect.~\ref{sect:ML_test} for details). We stress that they only provide a lower limit for 
the uncertainties in the determined abundances since they do not account for various 
systematic uncertainties that are unavoidable in the abundance analysis procedure. 

In the case of Li, the procedure was similar except for the following two points:

\begin{list}{--}{}
        \item
        Li abundances in \citet{DKB14} were determined by using the \textit{A}(Li)--$W$ interpolation formula 
        from \citet{SBC10} and $W$ was determined by fitting synthetic line profiles. Therefore, the line profile 
        fitting errors were estimated by measuring RMS deviation between the observed and best-fit Gaussian profiles; 
        \item
        an error in the determined Li abundance stemming from the use of the interpolation formula was taken into 
        account by using an estimated uncertainty of $\pm 0.01$\,dex \citep[][]{SBC10}.
\end{list}

The determined uncertainties are provided in Table~\ref{tab_app:abund_err}.

\begin{table}[t!]
        \begin{center}
                \caption{Results of the maximum-likelihood testing of the intrinsic spread in the 
                     abundances of Li, O, Na, Mg, and K.
                \label{tab:ML_values}}
                \begin{tabular}{lccc}                   
                        \hline 
                        \hline \\[-2.0ex]
                        Element, $X_{\rm i}$ & $\langle [X_{\rm i}/\ion{Fe}]\rangle$  & $\sigma^{[X_{\rm i}/\ion{Fe}]}$ & $\sigma_{\rm int}^{[X_{\rm i}/\ion{Fe}]}$   \\ [0.5ex]
                             &    dex                  &   dex    & dex             \\
                        \hline
                        %\vspace{5mm}
                        Li        & $1.49\pm0.02$  & $0.18$ & $0.15\pm0.02$  \\
                        O         & $0.33\pm0.01$  & $0.16$ & $0.12\pm0.01$  \\
                        Na        & $0.03\pm0.01$  & $0.14$ & $0.11\pm0.01$  \\
                        Mg        & $0.47\pm0.02$  & $0.12$ & $0.09\pm0.01$  \\
                        K         & $0.39\pm0.02$  & $0.09$ & $0.00\pm0.02$  \\
                        \hline
                \end{tabular}
        \end{center}
        \vspace{-5mm}
\end{table}

%=====================================================================
\subsubsection{Maximum-likelihood testing of the intrinsic spread in elemental abundances}\label{sect:ML_test}

In order to estimate the size of the possible intrinsic spread in the 1D~NLTE abundances of 
Li, O, Na, Mg, and K, we followed the procedure used in Paper~I, which is based on the prescription 
of \citet{MBI12,MBM15} that the authors applied to study the K abundance spreads in NGC~2419 and 
NGC~2808, and, later, also in 47~Tuc \citep{MMB17}. In the present work, the maximum-likelihood 
(ML) technique was utilized to evaluate the mean abundance ratio, $\xtmean{\mbox{[A/B]}}$, of 
elements A and B, as well as intrinsic spread, $\sigma_{\rm int}$, in the determined [A/B] abundance ratio. Here, we used [Li/Fe], 
[O/Fe], [Na/Fe], [Mg/Fe], and [K/Fe] abundance ratios in the TO stars, with the former three 
taken from \citet{DKB14} and the latter two determined in this study.
The obtained mean abundance ratio, $\langle [X_{\rm i}/\ion{Fe}]\rangle$, and 
its~uncertainty, as well as the total dispersion due to star-to-star abundance spread, $\sigma^{[X_{\rm i}/\ion{Fe}]}$, 
and the determined intrinsic abundance variation, $\sigma_{\rm int}^{[X_{\rm i}/\ion{Fe}]}$, 
in the abundance of element $X_{\rm i}$ are provided in Table~\ref{tab:ML_values}.

%%%%%%%%%%%%%%%%%%%%%%%%%%%%%%%%%%%%%%%%%%%%%%%%%%%%%%
\section{Results and discussion \label{sect:discuss}}
%%%%%%%%%%%%%%%%%%%%%%%%%%%%%%%%%%%%%%%%%%%%%%%%%%%%%%

%=====================================================================
\subsection{Average abundances and intrinsic abundance spreads in 47~Tuc \label{sect:discuss_abund}}

To our knowledge, the only studies of K abundance in 47~Tuc that have been carried out until now are 
those by \citet{CGB13} and \citet{MMB17}. The average 1D~NLTE potassium-to-iron abundance ratios obtained by \citet{CGB13} in three TO 
and nine SGB stars were $\langle{\rm [K/Fe]}\rangle_{\rm TO}=0.19\pm0.07$ and $\langle{\rm [K/Fe]}\rangle_{\rm SGB}=0.12\pm0.12$, 
respectively (the error is RMS deviation due to star-to-star abundance  variation). These values are 
compatible with the average 1D~NLTE abundance ratios obtained using RGB stars in Paper~I, $\langle{\rm[K/Fe]}\rangle_{\rm RGB}= 0.05 \pm0.13$. However, the average potassium-to-iron abundance ratio determined using TO stars in the present study, $\langle{\rm[K/Fe]}\rangle_{\rm TO}= 0.39 \pm0.09$, is 0.2\,dex higher than that obtained using three TO stars by \citet{CGB13}. We found that two stars are common to both samples. For them, the average abundance ratios obtained by \citet{CGB13} and determined in our study are $\langle{\rm[K/Fe]}\rangle_{\rm TO}= 0.18$ and $\langle{\rm[K/Fe]}\rangle_{\rm TO}= 0.32$ (the 1D~NLTE--LTE abundance corrections and microturbulence velocities used for these stars in the two studies are nearly identical). When corrected for the difference in iron abundance used by \citet[][$\feoh=-0.65$]{CGB13} and us (--0.76), the two values become nearly identical, with our [K/Fe] ratio being only 0.03\,dex higher. We therefore conclude that the difference in [K/Fe] ratios obtained in the TO samples by \citet{CGB13} and us is mostly due to the different [Fe/H] values used to compute [K/Fe] ratios. In addition, a substantial difference in the sample sizes may also lead to slightly different average [K/Fe] ratios.

The sample-averaged K abundance obtained in the study of 144 RGB stars in 47~Tuc by \citet{MMB17}, $\langle{\rm [K/Fe]}\rangle^{\rm 1D~NLTE} = -0.12 \pm 0.08$, is somewhat lower than that determined in Paper~I, $\langle{\rm[K/Fe]}\rangle_{\rm RGB}= 0.05 \pm0.13$. 
This difference may be a result of the different microturbulent velocties used in the two studies: the sample-averaged value in \citet{MMB17} is $\xi_{\rm t}=1.66$\,km/s while in our analysis we used 1.5\,km/s. Our tests show that the difference of 0.16\,km/s in $\xi_{\rm t}$ would lead to an $\sim 0.13$\,dex decrease in the average [K/Fe] ratio determined in Paper~I using RGB stars. With this taken into account, abundances obtained in the two studies would become very similar.
        
The origin of the significant difference between the average [K/Fe] ratios obtained by us in the TO and RGB stars, $0.34$\,dex, is not entirely clear, however. One possibility is that the value of microturbulent velocity used in our analysis of TO stars was in fact too low. We checked this using the six TO stars mentioned in Sect.~\ref{sect:abund_err} where we determined their iron abundances and microturbulence velocities using individual \ion{Fe}{i} lines. The average microturbulence velocity obtained in this way for the six stars was $\xi_{\rm t}=1.22\pm0.07$\,km/s, where error is RMS star-to-star variation. This value is significantly higher than the one used in the present study, 1.0\,km/s. With the higher value of $\xi_{\rm t}$, the average abundance obtained in our sample of TO stars would become $\approx0.1$\,dex lower. Still, this would still leave a difference of $\approx0.25$\,dex between the values obtained using TO and RGB stars.
On the other hand, it may also be that the microturbulence velocity used in our analysis of RGB stars, 1.5\,km/s, was slightly too low. For example, the $\xi_{\rm t}-\log g$ calibration of \citet{CGB09} that was also used in \citet{MMB17} would predict the average microturbulence velocity of $1.66$\,km/s for the gravity range of our RGB stars. 
Unfortunately, we could not obtain a reliable constraint on $\xi_{\rm t}$ in our RGB stars using spectroscopic means, due to an insufficient number of iron lines available in their spectra \citep[see][]{CKK17}. Nevertheless, the increase in $\xi_{\rm t}$ in both TO and RGB star samples would reduce the average abundance but the difference between two samples would remain almost the same.

Nevertheless, analysis of the six TO stars mentioned above revealed that star-to-star scatter in the determined microturbulence velocities was $\approx \pm 0.07$\,km/s. In terms of the determined K abundances, this would lead to a scatter of $\approx0.03$\,dex. Such star-to-star variation should have no detectable effect on various possible (anti-)correlations between the light element abundances. Therefore, our conclusions obtained in Sect.~\ref{sect:discuss_relat} below should remain unaffected.
 
The intrinsic abundance spreads of Li, O, Na, Mg, and K determined in our analysis are provided in Table~\ref{tab:ML_values}. In the case of K, we find zero intrinsic spread, $\sigma_{\rm int}^{\rm[K/Fe]}= 0.00\pm0.03$, identical to what was determined in Paper~I using RGB stars, $\sigma_{\rm int}^{\rm[K/Fe]}= 0.00\pm0.05$, and obtained by \citet{MMB17}, $\sigma_{\rm int}^{\rm[K/Fe]}= 0.00\pm0.02$. While the intrinsic spread of [Mg/Fe] in TO stars is similar to that determined by us in RGB stars, $\sigma_{\rm int}^{\rm[Mg/Fe]}= 0.08 \pm 0.02$ (Paper~I), in the present work we obtain considerably larger intrinsic spread in [Na/Fe], $\sigma_{\rm int}^{\rm[Na/Fe]}= 0.12 \pm 0.01$ (TO) versus $\sigma_{\rm int}^{\rm[Na/Fe]}= 0.04\pm0.05$ (RGB, Paper~I). The latter difference may be due to the lower quality of the RGB spectra, which led to larger abundance errors obtained using RGB stars and, thus, smaller intrinsic abundance spread. An intrinsic scatter of similar size was obtained in the case of Li and O for TO stars, $\sigma_{\rm int}^{\rm[Li/Fe]}= 0.14\pm0.02$ and $\sigma_{\rm int}^{\rm[O/Fe]}= 0.10\pm0.02$.

%=================================================================================
\subsection{Relations between the abundances of light elements and evolutionary properties of TO stars in 47~Tuc \label{sect:discuss_relat}}

As in our previous study, Paper~I, we used Student's $t$-test to verify the validity 
of the null hypothesis, that is, that the Pearson's correlation coefficient is equal to zero and, thus, there 
is no correlation in the abundance--iron and/or abundance--abundance planes (for simplicity, hereafter we will refer to the abundance ratios as abundances; we note that all abundance-to-iron ratios were obtained by scaling abundances of individual elements by the same constant value of [Fe/H]). For this, using each $x-y$ dataset 
shown in the panels of Fig.~\ref{fig:abund-ratios}, we computed the two-tailed probability, $p$, that the 
$t$-value in the given dataset could be equal or higher than its attained value when there is no correlation 
in the given $x-y$ plane. In all panels Pearson's correlation coefficients were computed by taking errors 
on both $x$ and $y$ axes into account. The obtained $p$-values are listed in Table~\ref{tab:pvalues}.

Our results suggest that there is no statistically significant relations in the [K/Fe]--[O/Fe] ($p=0.99$) 
and [K/Fe]--[Na/Fe] ($p=0.99$) planes. Similarly, we find no evidence for statistically significant relations 
in the [K/Fe]--[Li/Fe] ($p=0.99$), [K/Fe]--[Mg/Fe] ($p=0.99$), and [K/Fe]--[Na/O] ($p=0.99$) planes. 
We only found statistically significant relation in the [Mg/Fe]--[Na/Fe] plane ($p<0.0001$), which may suggest that the null hypothesis 
can be formally rejected on a high significance level (see below, however).\footnote{To test whether adding 
3D--1D abundance corrections may change our conclusions regarding the possible relations in different 
abundance-abundance planes, we also computed Student's $t$-values using 3D+NLTE abundances instead of those 
determined in 1D~NLTE. In all planes involving different abundance ratios, the obtained $p$-values were only 
slightly different from those determined earlier, thereby confirming the findings obtained in the 1D~NLTE case.}

In Fig.~\ref{fig:abund-radial} we show the determined [Mg/Fe] and [K/Fe] abundance ratios plotted versus the 
normalized distance from the cluster center, \textit{r}/\textit{r}$\rm_{h}$, were \textit{r} is the 
projected distance from the cluster center and \textit{r}$\rm_{h}$ is the half-light radius of 
47~Tuc taken from \citet[][$r_{\rm h}=174^{\prime\prime}$]{TDK93}. In both planes the obtained probabilities 
are $p\geq 0.55$ indicating that there are no statistically significant relations between the two abundance 
ratios and the projected distance from the cluster center.

Following \citet{KDB14} and Paper~I, we also investigated whether there are any significant 
relations between the kinematical properties of TO stars and the abundances of Mg and K in their 
atmospheres. For this we used absolute radial velocities of TO stars computed in \citet{KDB14}, 
 $\left|\Delta v_{\rm r}\right| \equiv \left| v_{\rm rad} - {\langle v_{\rm rad} \rangle}^{\rm clust} \right|$, 
where $v_{\rm rad}$ is the radial velocity of the individual star and ${\langle v_{\rm rad} \rangle}^{\rm clust}= -18.6$\,km/s 
is the mean radial velocity of the sample. As the $p$ -values of the $t$-test indicate (see Table \ref{tab:pvalues}), 
there are no statistically significant relations between the abundances of Mg and K and radial velocities.

These findings support our earlier results obtained in the analysis of Na, Mg, and K abundances in 32 RGB stars 
in Paper~I, where we found no statistically significant relations in the abundance-abundance, abundance-distance, 
and abundance-absolute radial velocity planes. On the other hand, a study of 144 RGB stars in 47~Tuc by \citet{MMB17} 
revealed statistically significant $\rm [K/Fe]-[Na/Fe]$ correlation and $\rm [K/Fe]-[O/Fe]$ anti-correlation 
(in the same study, such correlations were also detected in the globular cluster NGC 6752). In their analysis, 
\citet{MMB17} used Spearman's non-parametric rank-order correlation coefficients, $r_{\rm s}$, and computed the 
two-tailed probability, $p$, that in a given dataset $r_{\rm s}$ could attain a value that is equal to or larger 
than its measured value.

We therefore also performed non-parametric Spearman's 
and Kendall's $\tau$ rank-correlation tests using the data in our Figs.~\ref{fig:abund-ratios}-\ref{fig:disper-radial}. We also computed $p$-values using Pearson's correlation coefficients calculated without taking abundance errors into 
account. The $p$-values determined in all tests are provided in Table~\ref{tab:pvalues} and in the corresponding panels 
of Figs.~\ref{fig:abund-ratios}-\ref{fig:disper-radial}.

Except for the [Mg/Fe]--[Na/Fe] plane, the $p$-values computed using the Pearson's, Spearman's, and Kendall's 
correlation coefficients are all very similar but, at the same time, are significantly smaller than the $p$-values 
obtained using Pearson's coefficients computed with errors. In two planes, [K/Fe]--[Mg/Fe] and [K/Fe]--[Na/Fe], 
the obtained $p$-values are now sufficiently small to indicate the possible existence of weak correlations. In the 
[K/Fe]--[Mg/Fe] plane, however, the result may be influenced by the three stars with highest K abundances (for one of them K abundance and for two of them Na abundances were poorly 
determined as they were obtained from lines of quality classes B-C). When these points were removed from the 
analysis, Spearman's and Kendall's $p$-values became significantly larger, $p=0.22$ and $0.21$, respectively. 
Therefore, despite the relatively small $p$-values obtained in this plane using all data, we cannot reject with 
certainty the possibility that these small values are in fact a spurious result. In the [Mg/Fe]--[Na/Fe] plane, 
the $p$-values obtained in all three additional tests are significantly larger than the one determined by taking 
abundance errors into account. This may indicate that the very small $p$-value obtained by us in the analysis, when errors on both the $x$ and $y$ axes were taken into account, was spurious.
Therefore, we conclude that also in this case the null hypothesis cannot be rejected. Finally, no statistically 
significant relations were detected between abundances (see Figs.~\ref{fig:abund-radial}-\ref{fig:disper-radial}).

The possible existence of the weak correlation in the [K/Fe]--[Na/Fe] plane may be seen as being compatible 
with the result obtained by \citet{MMB17} who detected a correlation in the [K/Fe]--[Na/Fe] plane with Spearman's 
$p=0.017$. Although our Spearman's and Kendall's $p$-values are larger than those computed by \citet{MMB17}, the 
difference at least in part may be due to different sample sizes used in the two studies. Nevertheless, the null 
hypothesis, that there is no correlation between the two abundance ratios, cannot be rejected with confidence 
based alone on the $p$-values obtained in our Spearman and Kendall tests. We note that the analysis performed on subsamples of stars selected according to the quality class of spectral lines did not reveal any significant relations in any of the data planes studied above.

\begin{table}[t!]
        \begin{center}
        \caption{Pearson's, Spearman's, and Kendall's two-tailed $p$-values for various abundance-abundance, 
                 abundance-$r/r_{\rm h}$, and $\rm\left|\Delta v_{\rm r}\right|$-abundance velocity planes.}
                \resizebox{\columnwidth}{!}{            
                \begin{tabular}{lr@{}lr@{}lr@{}lr@{}l}
                        \hline\hline
                        Plane & \multicolumn{2}{c}{Pearson} & \multicolumn{2}{c}{Pearson} & \multicolumn{2}{c}{Spearman} & \multicolumn{2}{c}{Kendall} \\ [0.5ex]
                         &&  $p$-value$\tablefootmark{1}$  && $p$-value$\tablefootmark{2}$     &&    $p$-value     &&  $p$-value    \\
                        \hline
            $\rm[K/Fe]-[Li/Fe]$                            &  &$0.999$  &  &$0.929$   &  &$0.824$  &  &$0.781$   \\                       
            $\rm[K/Fe]-[O/Fe]$                             &  &$0.999$  &  &$0.296$   &  &$0.557$  &  &$0.500$   \\
            $\rm[K/Fe]-[Na/Fe]$                            &  &$0.999$  &  &$0.050$   &  &$0.076$  &  &$0.060$   \\
            $\rm[K/Fe]-[Mg/Fe]$                            &  &$0.995$  &  &$0.006$   &  &$0.024$  &  &$0.022$   \\                       
            $\rm[Mg/Fe]-[Na/Fe]$                           & <&$0.0001$ &  &$0.220$   &  &$0.288$  &  &$0.258$   \\
            $\rm[K/Fe]-[Na/O]$                             &  &$0.999$  &  &$0.115$   &  &$0.231$  &  &$0.175$   \\
            $\rm[Mg/Fe]-r/r_{\rm h}$                       &  &         &  &$0.995$   &  &$0.741$  &  &$0.729$   \\
            $\rm[K/Fe]-r/r_{\rm h}$                        &  &         &  &$0.567$   &  &$0.578$  &  &$0.545$   \\
            $\rm\left|\Delta v_{\rm r}\right|-[Mg/Fe]$     &  &         &  &$0.722$   &  &$0.743$  &  &$0.770$   \\
            $\rm\left|\Delta v_{\rm r}\right|-[K/Fe]$      &  &         &  &$0.655$   &  &$0.626$  &  &$0.657$   \\ 
                        \hline
                \end{tabular}}
                \tablefoot{
                \tablefoottext{1}{Taking into account errors on both $x-y$ axes.}
                \tablefoottext{2}{Without $x-y$ errors.}
                }
            \label{tab:pvalues}
        \end{center}
\end{table}

%%%%%%%%%%%%%%%%%%%%%%%%%%%%%%%%%%%%%%%%
\section{Conclusions}\label{sect:conclus}
%%%%%%%%%%%%%%%%%%%%%%%%%%%%%%%%%%%%%%%%

We performed abundance analyses of Mg and K in the turn-off stars of the Galactic 
globular cluster 47~Tuc. Abundances were determined using archival VLT~FLAMES/GIRAFFE 
spectra that were obtained in HR~18 setup ($746.8-788.9$\,nm, $\textit{R}=18\,400$). 
Spectroscopic data were analyzed using 1D \ATLAS\ model atmospheres and 1D~NLTE abundance 
analysis methodology. One-dimensional~NLTE spectral line synthesis was performed with the \MULTI\ 
package, using up-to-date model atoms of Mg and K. We also used 3D hydrodynamical \COBOLD\ 
and 1D hydrostatic \LHD\ model atmospheres to compute 3D--1D abundance corrections for 
the spectral lines of \ion{Mg}{i} and \ion{K}{i} utilized in this study. The obtained abundance 
corrections were small, in all cases $\textless0.1$\,dex, indicating that the influence of 
convection on the formation of these spectral lines in the atmospheres of TO stars in 47~Tuc should be minor. 

The determined sample-averaged abundance ratios are $\langle{\rm[Mg/Fe]}\rangle^{\rm 1D~NLTE} =0.47\pm0.12$ 
and $\langle{\rm[K/Fe]}\rangle^{\rm 1D~NLTE}=0.39\pm0.09$ (numbers after 
the $\pm$ sign are RMS abundance variations due to star-to-star scatter). In the case of Mg we 
find small but significant intrinsic star-to-star scatter in the [Mg/Fe] abundance ratio, 
$\sigma_{\rm int}^{[{\rm Mg}/\ion{Fe}]}=0.10\pm0.01$. No intrinsic scatter was found for K. 
Both results are in line with our earlier findings obtained using RGB stars in 47~Tuc (Paper~I). 
Abundances of another three light elements, Li, O, and Na, that were determined in our sample stars 
earlier by \citet{DKB14}, also show intrinsic scatter on levels similar to that of Mg, $\sigma_{\rm int}^{[{\rm Li,O,Na}/\ion{Fe}]}=0.10$ to $0.12$\,dex.

Although our data suggest the existence of a weak correlation in the [K/Fe]--[Na/Fe] plane, its 
statistical significance is not high enough to claim its existence with confidence. 
We also detected no statistically significant correlations or anti-correlations between [Mg/Fe] 
and [K/Fe] abundance ratios and projected distance from the cluster center. Finally, we found 
no relations between the absolute radial velocities of individual stars and abundances of Mg 
and K in their atmospheres. 

The absence of statistically significant relations between abundances of different light elements, as well as those between their abundances and the kinematical properties of the host stars, is in good agreement with the results obtained in our previous analysis of RGB 
stars in 47~Tuc (Paper I). The results of the present analysis may be seen to support the 
existence of a possible relation in the [K/Fe]--[Na/Fe] plane, as found in \citealt{MMB17}, 
albeit at a lower level of significance. However, we find no statistically significant relation in 
the [K/Fe]--[O/Fe] plane. Differences in the statistical significance levels obtained in the two 
studies may in part be due to the factor of two difference in the sample sizes used. 
Therefore, further homogeneous analyses of larger stellar samples using higher quality spectroscopic 
data may be needed to shed further light on this controversial issue.

%=================================================================================
\begin{acknowledgements}
%=================================================================================

We would like to thank the anonymous referee for her or his constructive comments and 
suggestions that have helped to improve the paper.
This work was supported by grants from the Research Council of Lithuania (MIP-089/2015, 
TAP LZ 06/2013). The study was based on observations made with the European Southern 
Observatory telescopes obtained from the ESO/ST-ECF Science Archive Facility.

\end{acknowledgements}

%________________________________________________________________

\bibliographystyle{aa}

\

%\begin{comment}
\begin{appendix}

%%%%%%%%%%%%%%%%%%%%%%%%%%%%%%%%%%%%%%%%
\section{Determination of Mg and K abundances in the Sun and Arcturus \label{sect_app:ArcSun}}
%%%%%%%%%%%%%%%%%%%%%%%%%%%%%%%%%%%%%%%%

To verify the adopted model atoms and atomic parameters of \ion{Mg}{i} and \ion{K}{i} lines that were 
used in our study of TO stars in 47~Tuc, we used the \MULTI\ code to compute their synthetic 1D~NLTE line 
profiles and determined abundances of these elements in the Sun and Arcturus. The obtained solar abundances were 
also used to compute element-to-iron abundance ratios in the sample of TO stars (Sect.~\ref{sect:MgK47Tuc}.

%=====================================================================
\subsection{Observed spectra and abundance determination}
%=====================================================================

For Mg and K 1D~NLTE abundance determination in the Sun, we used the re-reduced Kitt Peak Solar 
Flux atlas from \citet{K06}. The spectrum covers a range of 300--1000\,nm, with $\textit{R}=523\,000$, 
and $S/N \sim 4000$ in the infrared part of the spectrum. A solar model atmosphere computed with the
\ATLAS\ code was used in the abundance determination. The model atmosphere was computed using
$\Teff = 5777 \pm 10$\,K and $\logg=4.43 \pm 0.02$ \citep{ASD16}.
The value of solar microturbulence velocity used in the 
abundance determination, $\xi_{\rm t}=1.01 \pm 0.06$\,km/s, was also taken from \citet[][]{ASD16}.

\begin{figure}[t]
        \begin{center}
                \includegraphics[scale=0.90]{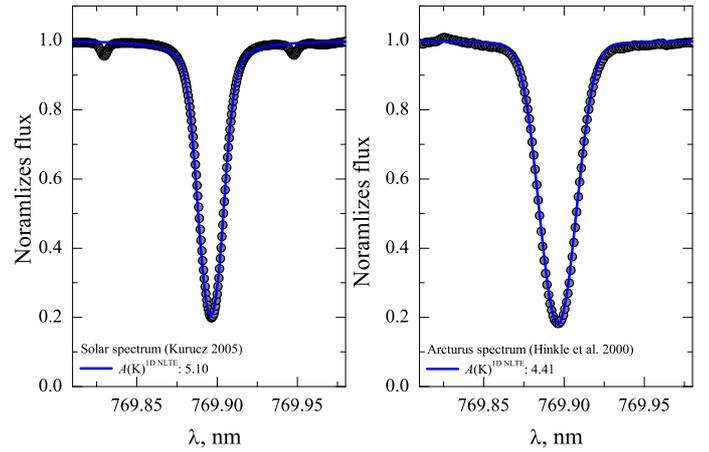}
                \caption{Potassium line profiles in the spectra of the Sun and Arcturus. 
                         Gray dots show the observed spectra and the blue solid lines are                        theoretical 1D~NLTE line profiles computed using the \MULTI\ code.}
                \label{appfig:K_ArcSun}
        \end{center}
\end{figure}

\begin{figure}[t]
        \begin{center}
                \includegraphics[scale=0.90]{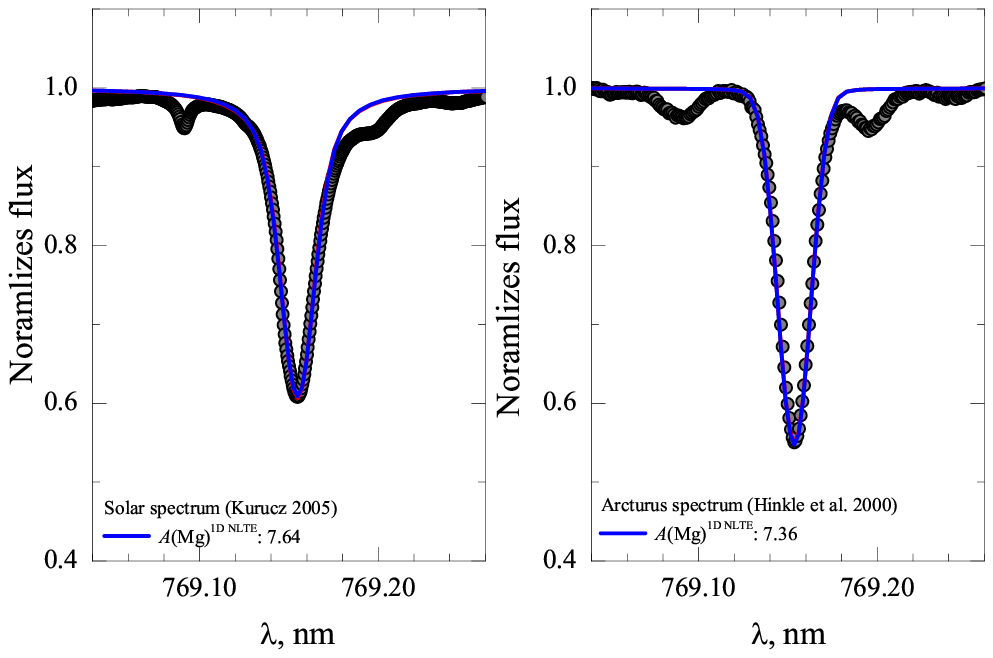}
                \caption{Magnesium line profiles in the spectra of the Sun and Arcturus. 
                                 Gray dots show the observed spectra and the blue solid lines are 
                         theoretical 1D~NLTE line profiles computed using the \MULTI\ code.}
                \label{appfig:Mg_ArcSun}
        \end{center}
\end{figure}

In the case of Arcturus, we used a spectral atlas from \citet{HWV00}. The spectrum covers a wavelength 
range of 372.7--930.0\,nm, has a resolution of $\textit{R}=150\,000$, and  $S/N \sim 1000$ in 
the near-infrared part of the spectrum. For the abundance determination we computed an ATLAS9 model 
atmosphere using the atmospheric parameters from \citet{RA11}: $\Teff = 4286 \pm 30$\,K, $\logg=1.66 \pm 0.05$, 
and $\FeoH=-0.52 \pm 0.04$. The value of microturbulent velocity, $\xi_{\rm t}=1.58 \pm 0.12$\,km/s, was taken from \citet{JHS15}.
The fits of Mg and K line profiles in the spectra of the Sun and Arcturus are shown in 
Figs.~\ref{appfig:K_ArcSun} and~\ref{appfig:Mg_ArcSun}.

%=====================================================================
\subsection{Errors in the determined abundances of Mg and K}
%=====================================================================

To estimate errors in the determined K and Mg abundances in the Sun and Arcturus, we used the same 
methodology that was employed to obtain abundance determination errors for TO stars 
(Sect.~\ref{sect:abund_err}). The determined individual errors arising due to inaccuracies in the determined effective 
temperature, gravity, microturbulence velocity, continuum placement, and the line profile fit, as well as 
the total error in the determined abundance, are provided in columns 3-8 of 
Table~\ref{tab_app:abund_err_Arcturus+Sun}.

\begin{table}[t!]
        \begin{center}
                \caption{Errors in the determined abundances of Mg and K in the atmospheres 
                     of the Sun and Arcturus, $\sigma(A)_{\rm tot}$, obtained by changing various atmospheric parameters by their errors. The sign $\pm$ or $\mp$ reflects the change 
                     in the elemental abundance, which occurs due to an increase (top sign) 
                     or decrease (bottom sign) in a given atmospheric parameter.}
                \vspace{-5mm}
                \resizebox{\columnwidth}{!}{            
                        \begin{tabular}{lccccccccc}
                                \hline\hline
                                Element & Line & $\sigma(\Teff)$ & $\sigma(\log g)$  &      $\sigma(\xi_{\rm t})$    &  $\sigma(\rm cont)$ &  $\sigma(\rm fit)$  & $\sigma(A)_{\rm tot}$  \\ [0.5ex]
                                & $\lambda$, nm          &       dex              &             dex           &        dex         &   dex    &    dex    &   dex                         \\
                                \hline
                                \textbf{Sun}     & & & & & & & &                                                                            \\
                                \ion{Mg}{i}      & 769.16 & $\pm 0.004$ & $\mp 0.001$  & $\mp 0.001$ & $\pm 0.009$  & $\pm 0.025$ &  0.026  \\
                                \ion{K}{i}       & 769.16 & $\pm 0.009$ & $\mp 0.005$  & $\mp 0.009$ & $\pm 0.008$  & $\pm 0.055$ &  0.057  \\
                                \textbf{Arcturus} & & & & & & & &                                                                           \\
                                \ion{Mg}{i}      & 769.16 & $\pm 0.009$ & $\mp 0.012$  & $\mp 0.015$ & $\pm 0.022$  & $\pm 0.045$ &  0.054      \\
                                \ion{K}{i}       & 769.89 & $\pm 0.033$ & $\mp 0.001$  & $\mp 0.088$ & $\pm 0.028$  & $\pm 0.059$ &  0.114       \\
                                \hline
                        \end{tabular}}
                        \label{tab_app:abund_err_Arcturus+Sun}
                \end{center}
                
                \vspace{-5mm}
        \end{table}

%=====================================================================
\subsection{Abundances of Mg and K in the Sun and Arcturus}
%=====================================================================

The determined solar 1D~NLTE abundance of potassium is $\textit{A}(\rm K)^{1D~NLTE}_{\odot}=5.10 \pm 0.06$. 
The obtained 1D~LTE potassium abundance is $\textit{A}(\rm K)^{1D~LTE}_{\odot}=5.20 \pm 0.06$, 
which is slightly lower than $\textit{A}(\rm K)^{1D~LTE}_{\odot}=5.31 \pm 0.02$ obtained by \citet{RA11}. In the case of 
magnesium, we determined $\textit{A}(\rm Mg)^{1D~NLTE}_{\odot}=7.64 \pm 0.03$ and $\textit{A}(\rm Mg)^{1D~LTE}_{\odot}=7.67 \pm 0.03$. 
The latter value is slightly larger than $\textit{A}(\rm K)^{1D~LTE}_{\odot}=7.59 \pm 0.02$ determined by \citet{RA11}. 

For Arcturus, we determined $\textit{A}(\rm K)^{1D~NLTE}=4.41 \pm 0.11$. 
The 1D~LTE potassium abundance, $\textit{A}(\rm K)^{1D~LTE}=5.01 \pm 0.11$, agrees very well with 
$\textit{A}(\rm K)^{1D~LTE}=4.99 \pm 0.07$ obtained by \citet{RA11}. For magnesium, we obtained 
$\textit{A}(\rm Mg)^{1D~NLTE}=7.36 \pm 0.05$. In the case of 1D~LTE abundances, again, we obtained a good 
agreement between our 1D~LTE abundance, $\textit{A}(\rm Mg)^{1D~LTE}=7.46 \pm 0.05$, 
and $\textit{A}(\rm Mg)=7.47 \pm 0.03$ from \citet{RA11}.

%%%%%%%%%%%%%%%%%%%%%%%%%%%%%%%%%%%%%%%%
\section{Abundances of Mg and K in the atmospheres of TO stars of 47~Tuc \label{sect_app:abund}}
%%%%%%%%%%%%%%%%%%%%%%%%%%%%%%%%%%%%%%%%

The 1D~NLTE [Mg/Fe] and [K/Fe] abundance ratios determined in the sample of TO stars in 47~Tuc are provided in Table~\ref{tab_app:all_abund}. A detailed description of the methodology used to determine elemental abundances is given in Sect.~\ref{sect:1Dabund}. The contents of Table~\ref{tab_app:all_abund} are:

\begin{itemize}
        \item column~1: object identification (ID);
        \item column~2: right ascension;
        \item column~3: declination;
        \item column~4: effective temperature;
        \item column~5: surface gravity;
        \item column~6: radial velocity;
        \item column~7: 1D NLTE magnesium abundance and its error;
        \item column~8: 1D NLTE potassium abundance and its error.
\end{itemize}

\noindent The sample-averaged values of the determined abundance ratios and their RMS values are provided in the last two lines of Table~\ref{tab_app:all_abund}.

\onecolumn

\begin{center}
\begin{longtable}{lccccr@{}lcc}
\caption{Target TO stars in 47~Tuc, their atmospheric parameters, and determined [Mg/Fe] and [K/Fe] abundance ratios.}
\label{tab_app:all_abund}\\                                   
\hline\hline
\smallskip  
Star         & RA       & Dec.   & \Teff   & $\log g$ &  \multicolumn{2}{c}{$v_{\rm rad}$} &   ${\rm[Mg/Fe]}$   & ${\rm[K/Fe]}$  \\
ID         & (2000)   & (2000) &  K      & [cgs]    &   km  &   /s                       &    1D NLTE         & 1D NLTE        \\
\hline      
\vspace{-0.2 cm}
\endfirsthead
\hline      
\vspace{-0.2 cm}
\endhead    
\hline      
\endfoot    
00006129  & 6.15846 & --71.96322 & 5851 & 4.06 & --21&.1 &  0.42 $\pm$0.07 &  0.33 $\pm$0.13  \\
00006340  & 5.96746 & --71.96075 & 5817 & 4.02 &  --7&.4 &  0.34 $\pm$0.07 &  0.37 $\pm$0.13  \\
00007619  & 6.33533 & --71.94289 & 5790 & 4.05 & --10&.9 &  ...            &  0.41 $\pm$0.13  \\
00007969  & 6.11763 & --71.93814 & 5811 & 4.06 & --28&.5 &  ...            &  0.45 $\pm$0.13  \\
00008359  & 6.24488 & --71.93133 & 5839 & 4.06 & --23&.0 &  0.62 $\pm$0.07 &  0.60 $\pm$0.13  \\
00008881  & 6.16508 & --71.92217 & 5916 & 4.10 & --17&.1 &  0.45 $\pm$0.07 &  ...             \\
00009191  & 6.21133 & --71.91592 & 5826 & 4.07 & --26&.7 &  0.54 $\pm$0.07 &  0.43 $\pm$0.13  \\
00009243  & 6.27892 & --71.91464 & 5857 & 4.08 & --27&.9 &  0.64 $\pm$0.07 &  0.38 $\pm$0.13  \\
00009434  & 6.24204 & --71.91056 & 5872 & 4.08 & --21&.8 &  ...            &  0.36 $\pm$0.13  \\
00009540  & 6.11050 & --71.90853 & 5843 & 4.06 &  --7&.1 &  ...            &  0.38 $\pm$0.13  \\
00014912  & 5.80258 & --71.96294 & 5859 & 4.08 & --12&.9 &  0.65 $\pm$0.07 &  0.33 $\pm$0.13  \\
00015086  & 5.76054 & --71.96000 & 5878 & 4.07 & --30&.8 &  ...            &  0.44 $\pm$0.13  \\
00015174  & 5.82779 & --71.95847 & 5788 & 4.04 & --24&.3 &  ...            &  ...             \\
00015346  & 5.58437 & --71.95531 & 5725 & 4.03 & --10&.5 &  ...            &  0.39 $\pm$0.13  \\
00016131  & 5.77725 & --71.94094 & 5823 & 4.07 & --10&.2 &  0.46 $\pm$0.07 &  0.44 $\pm$0.13  \\
00016631  & 5.75729 & --71.92917 & 5820 & 4.08 & --27&.2 &  0.68 $\pm$0.07 &  ...             \\
00017628  & 5.87896 & --71.90236 & 5925 & 4.11 & --21&.1 &  0.39 $\pm$0.07 &  0.36 $\pm$0.13  \\
00017767  & 5.84779 & --71.89828 & 5812 & 4.06 & --25&.9 &  0.53 $\pm$0.07 &  0.50 $\pm$0.13  \\
00031830  & 5.41504 & --72.04769 & 5832 & 4.05 & --18&.6 &  0.36 $\pm$0.07 &  0.29 $\pm$0.13  \\
00036086  & 5.70875 & --72.20400 & 5850 & 4.05 & --14&.3 &  ...            &  0.36 $\pm$0.13  \\
00036747  & 5.77333 & --72.19608 & 5814 & 4.09 &  --9&.8 &  ...            &  0.46 $\pm$0.13  \\
00038656  & 5.62004 & --72.17497 & 5850 & 4.08 & --19&.2 &  ...            &  0.30 $\pm$0.13  \\
00040049  & 5.74092 & --72.16181 & 5822 & 4.07 & --14&.2 &  0.41 $\pm$0.07 &  0.37 $\pm$0.13  \\
00040087  & 5.53888 & --72.16119 & 5787 & 4.03 & --27&.0 &  0.31 $\pm$0.07 &  0.31 $\pm$0.13  \\
00040355  & 5.72492 & --72.15906 & 5879 & 4.06 &  --7&.1 &  0.57 $\pm$0.07 &  0.45 $\pm$0.13  \\
00043095  & 5.67775 & --72.13700 & 5770 & 4.05 & --23&.4 &  0.49 $\pm$0.07 &  ...             \\
00043108  & 5.57883 & --72.13678 & 5797 & 4.03 & --19&.5 &  ...            &  ...             \\
00044983  & 5.71950 & --72.12375 & 5848 & 4.04 & --14&.3 &  0.56 $\pm$0.07 &  0.43 $\pm$0.13  \\
00045982  & 5.64500 & --72.11706 & 5707 & 4.00 & --16&.4 &  0.56 $\pm$0.07 &  0.46 $\pm$0.13  \\
00046498  & 5.51050 & --72.11339 & 5790 & 4.04 &  --7&.5 &  0.45 $\pm$0.07 &  0.28 $\pm$0.13  \\
00049829  & 5.76571 & --72.09175 & 5740 & 3.99 & --25&.9 &  0.54 $\pm$0.07 &  0.38 $\pm$0.13  \\
00051341  & 5.55921 & --72.08197 & 5731 & 4.01 &  --3&.2 &  0.50 $\pm$0.07 &  0.41 $\pm$0.13  \\
00051740  & 5.53704 & --72.07939 & 5857 & 4.07 & --28&.9 &  0.50 $\pm$0.07 &  0.51 $\pm$0.13  \\
00052108  & 5.50988 & --72.07694 & 5688 & 3.99 & --16&.0 &  0.36 $\pm$0.07 &  0.14 $\pm$0.13  \\
00054596  & 5.61767 & --72.06100 & 5825 & 4.05 & --16&.7 &  0.49 $\pm$0.07 &  0.37 $\pm$0.13  \\
00058492  & 5.68208 & --72.03306 & 5728 & 4.02 & --11&.1 &  ...            &  ...             \\
00059579  & 5.66825 & --72.02414 & 5660 & 3.98 &  --7&.9 &  ...            &  ...             \\
00061639  & 5.69313 & --72.00528 & 5779 & 4.03 & --20&.0 &  ...            &  ...             \\
00062314  & 5.56467 & --71.99794 & 5740 & 4.03 &   1&.3  &  ...            &  ...             \\
00062737  & 5.58004 & --71.99319 & 5691 & 4.01 & --20&.4 &  ...            &  ...             \\
00062773  & 5.87338 & --71.99317 & 5854 & 4.05 & --20&.0 &  0.47 $\pm$0.07 &  0.37 $\pm$0.13  \\
00063201  & 5.60025 & --71.98767 & 5759 & 4.02 & --13&.8 &  ...            &  ...             \\
00063954  & 5.77167 & --71.97908 & 5801 & 4.02 & --13&.2 &  ...            &  ...             \\
00063973  & 5.70850 & --71.97875 & 5780 & 4.02 &  --8&.1 &  ...            &  ...             \\
00065981  & 6.05225 & --72.22219 & 5814 & 4.07 & --26&.2 &  ...            &  ...             \\
00066603  & 6.05375 & --72.21225 & 5848 & 4.08 & --10&.9 &  ...            &  0.39 $\pm$0.13  \\
00066813  & 6.34237 & --72.20903 & 5817 & 4.07 & --16&.8 &  0.33 $\pm$0.07 &  0.39 $\pm$0.13  \\
00066840  & 6.25950 & --72.20878 & 5780 & 4.10 & --14&.2 &  ...            &  0.51 $\pm$0.13  \\
00067280  & 6.02708 & --72.20253 & 5808 & 4.07 &  --8&.2 &  ...            &  0.35 $\pm$0.13  \\
00069585  & 6.29904 & --72.17517 & 5888 & 4.08 & --13&.3 &  ...            &  0.31 $\pm$0.13  \\
00070686  & 6.22921 & --72.16494 & 5808 & 4.05 &  --8&.3 &  0.50 $\pm$0.07 &  0.46 $\pm$0.13  \\
00070910  & 6.27663 & --72.16297 & 5797 & 4.06 &  --7&.3 &  ...            &  0.30 $\pm$0.13  \\
00071404  & 6.29454 & --72.15886 & 5787 & 4.06 & --20&.1 &  0.55 $\pm$0.07 &  ...             \\
00072011  & 6.11733 & --72.15458 & 5702 & 4.05 & --12&.4 &  ...            &  ...             \\
00096225  & 6.27933 & --72.02936 & 5805 & 4.04 & --32&.0 &  ...            &  0.23 $\pm$0.13  \\
00097156  & 6.36075 & --72.02406 & 5750 & 4.03 & --15&.7 &  0.53 $\pm$0.07 &  ...             \\
00099636  & 6.26008 & --72.00881 & 5799 & 4.06 & --20&.2 &  0.53 $\pm$0.07 &  0.26 $\pm$0.13  \\
00100325  & 6.35675 & --72.00369 & 5794 & 4.05 & --18&.1 &  ...            &  0.42 $\pm$0.13  \\
00102294  & 6.06792 & --71.98808 & 5772 & 4.01 & --23&.0 &  0.40 $\pm$0.07 &  0.48 $\pm$0.13  \\
00102307  & 6.21471 & --71.98781 & 5835 & 4.07 & --25&.2 &  0.33 $\pm$0.07 &  0.44 $\pm$0.13  \\
00103067  & 5.94763 & --71.98056 & 5665 & 3.98 & --17&.7 &  ...            &  ...             \\
00103709  & 6.02521 & --71.97353 & 5806 & 4.02 & --22&.6 &  ...            &  ...             \\
00104049  & 6.17200 & --71.96964 & 5768 & 4.02 & --24&.2 &  0.65 $\pm$0.07 &  0.62 $\pm$0.13  \\
00106794  & 6.47321 & --72.18328 & 5789 & 4.08 & --13&.7 &  0.64 $\pm$0.07 &  0.48 $\pm$0.13  \\
00107260  & 6.45896 & --72.17064 & 5829 & 4.06 & --15&.7 &  ...            &  0.40 $\pm$0.13  \\
00107528  & 6.57650 & --72.16361 & 5923 & 4.11 & --10&.0 &  ...            &  ...             \\
00107618  & 6.59092 & --72.16119 & 5831 & 4.06 & --24&.4 &  0.34 $\pm$0.07 &  ...             \\
00107866  & 6.56246 & --72.15469 & 5727 & 4.06 &  --9&.1 &  ...            &  ...             \\
00108171  & 6.40738 & --72.14778 & 5812 & 4.06 &  --9&.1 &  ...            &  0.34 $\pm$0.13  \\
00108389  & 6.58104 & --72.14253 & 5808 & 4.05 & --18&.2 &  ...            &  0.47 $\pm$0.13  \\
00109441  & 6.53275 & --72.11814 & 5875 & 4.07 & --21&.8 &  ...            &  ...             \\
00109777  & 6.50933 & --72.11058 & 5873 & 4.09 & --19&.8 &  ...            &  0.50 $\pm$0.13  \\
00110197  & 6.60008 & --72.10139 & 5824 & 4.08 & --16&.6 &  ...            &  ...             \\
00111136  & 6.48775 & --72.08114 & 5908 & 4.08 & --29&.9 &  ...            &  0.35 $\pm$0.13  \\
00111231  & 6.52613 & --72.07919 & 5732 & 4.03 & --11&.5 &  ...            &  ...             \\
00112473  & 6.59492 & --72.05506 & 5840 & 4.05 & --17&.3 &  0.40 $\pm$0.07 &  0.21 $\pm$0.13  \\
00112684  & 6.46096 & --72.05136 & 5780 & 4.03 & --29&.1 &  ...            &  0.44 $\pm$0.13  \\
00113090  & 6.47025 & --72.04353 & 5794 & 4.05 & --15&.4 &  ...            &  0.44 $\pm$0.13  \\
00113841  & 6.55175 & --72.02800 & 5854 & 4.07 & --23&.5 &  ...            &  0.36 $\pm$0.13  \\
00113959  & 6.49396 & --72.02594 & 5968 & 4.10 & --27&.4 &  ...            &  0.49 $\pm$0.13  \\
00115880  & 6.51471 & --71.98331 & 5845 & 4.06 & --13&.7 &  ...            &  0.36 $\pm$0.13  \\
10000002  & 5.43304 & --72.05411 & 5894 & 4.09 &  --6&.8 &  ...            &  0.31 $\pm$0.13  \\
10000004  & 5.62229 & --72.10428 & 5934 & 4.10 & --22&.7 &  0.36 $\pm$0.07 &  0.28 $\pm$0.13  \\
10000008  & 5.70025 & --72.15828 & 5905 & 4.13 & --17&.9 &  ...            &  0.34 $\pm$0.13  \\
10000009  & 5.70075 & --72.09483 & 5792 & 4.09 &  --7&.2 &  ...            &  ...             \\
10000012  & 5.70475 & --72.08533 & 5836 & 4.03 & --19&.0 &  ...            &  ...             \\
10000015  & 5.72129 & --72.07636 & 5754 & 4.02 &  --8&.1 &  ...            &  ...             \\
10000016  & 5.72533 & --72.02817 & 5724 & 3.98 &  --6&.7 &  ...            &  ...             \\
10000020  & 5.75263 & --72.06483 & 5834 & 4.10 & --21&.9 &  0.36 $\pm$0.07 &  0.23 $\pm$0.13  \\
10000022  & 5.76167 & --72.04869 & 5749 & 3.99 & --17&.5 &  0.63 $\pm$0.07 &  ...             \\
10000026  & 5.77117 & --72.12517 & 5784 & 4.06 &  --9&.8 &  0.52 $\pm$0.07 &  0.20 $\pm$0.13  \\
10000027  & 5.77721 & --72.12919 & 5829 & 4.08 & --20&.4 &  0.40 $\pm$0.07 &  ...             \\
10000036  & 5.84604 & --72.00550 & 5706 & 4.00 & --25&.4 &  ...            &  ...             \\
10000038  & 5.86846 & --72.19789 & 5810 & 4.00 &  --5&.1 &  0.36 $\pm$0.07 &  0.35 $\pm$0.13  \\
10000041  & 5.90950 & --71.93806 & 5889 & 4.08 & --27&.4 &  ...            &  ...             \\
10000043  & 5.94513 & --72.17733 & 5883 & 4.05 & --20&.9 &  0.73 $\pm$0.07 &  0.54 $\pm$0.13  \\
10000048  & 5.99058 & --71.98381 & 5832 & 4.05 & --22&.2 &  ...            &  0.45 $\pm$0.13  \\
10000049  & 6.00479 & --72.18656 & 5935 & 4.14 & --15&.5 &  0.31 $\pm$0.07 &  ...             \\
10000053  & 6.04242 & --72.20942 & 5881 & 4.08 & --20&.0 &  0.41 $\pm$0.07 &  0.47 $\pm$0.13  \\
10000057  & 6.08746 & --71.93789 & 5846 & 4.10 & --18&.2 &  0.36 $\pm$0.07 &  0.45 $\pm$0.13  \\
10000062  & 6.12154 & --71.97469 & 5891 & 4.10 & --13&.3 &  ...            &  0.49 $\pm$0.13  \\
10000068  & 6.15775 & --71.95836 & 5923 & 4.09 & --21&.8 &  0.28 $\pm$0.07 &  0.49 $\pm$0.13  \\
10000072  & 6.19088 & --71.97972 & 5829 & 4.09 & --17&.1 &  0.50 $\pm$0.07 &  0.37 $\pm$0.13  \\
10000073  & 6.21196 & --72.00553 & 5855 & 4.03 & --24&.5 &  ...            &  ...             \\
10000075  & 6.24354 & --71.96136 & 5882 & 4.08 & --23&.4 &  0.31 $\pm$0.07 &  0.47 $\pm$0.13  \\
10000079  & 6.27275 & --72.12033 & 5750 & 4.09 & --29&.6 &  ...            &  0.41 $\pm$0.13  \\
10000086  & 6.30192 & --72.05958 & 5708 & 3.99 &  --7&.4 &  0.26 $\pm$0.07 &  0.30 $\pm$0.13  \\
10000088  & 6.31217 & --72.03944 & 5771 & 4.07 & --16&.0 &  0.43 $\pm$0.07 &  0.30 $\pm$0.13  \\
10000090  & 6.34033 & --71.96881 & 5921 & 4.08 & --15&.3 &  0.53 $\pm$0.07 &  0.34 $\pm$0.13  \\
10000094  & 6.42554 & --72.07425 & 5869 & 4.10 & --20&.8 &  0.37 $\pm$0.07 &  0.30 $\pm$0.13  \\
\hline                                                                                
Sample-average         &         &       &      &      &     &              & 0.47            & 0.39        \\
RMS                    &         &       &      &      &     &              & 0.12            & 0.09        \\                                                   

\end{longtable}                                          
\end{center}

\end{appendix}
\end{document}